\documentclass[journal=nalefd,manuscript=letter]{achemso}

\usepackage{chemformula} 
\usepackage[T1]{fontenc} 
\usepackage[colorlinks=false,bookmarks=false,citecolor=black,urlcolor=black]{hyperref} 
\usepackage{xcolor} 
\usepackage[separate-uncertainty=true]{siunitx} 
\usepackage{float} 
\usepackage{url}
\usepackage{hyphenat}
\usepackage{siunitx}

\author{Kirill Koshelev}
    \affiliation{Nonlinear Physics Center, Research School of Physics, Australian National University, Canberra ACT 2601, Australia}
    \email{ki.koshelev@gmail.com}
\author{Yutao Tang}
    \affiliation{Department of Materials Science and Engineering, Shenzhen Institute for Quantum Science and Engineering, Southern University of Science and Technology, Shenzhen 518055, China}
\author{Zixian Hu}
    \affiliation{Department of Materials Science and Engineering, Shenzhen Institute for Quantum Science and Engineering, Southern University of Science and Technology, Shenzhen 518055, China}
\author{Ivan I. Kravchenko}
    \affiliation{Center for Nanophase Materials Sciences, Oak Ridge National Laboratory, Oak Ridge, Tennessee 37831, United States}
\author{\\Guixin Li}
    \affiliation{Department of Materials Science and Engineering, Shenzhen Institute for Quantum Science and Engineering, Southern University of Science and Technology, Shenzhen 518055, China}
\author{Yuri Kivshar}
    \affiliation{Nonlinear Physics Center, Research School of Physics, Australian National University, Canberra ACT 2601, Australia}
    \email{yuri.kivshar@anu.edu.au}

\title{Resonant chiral effects  in nonlinear dielectric metasurfaces}


\begin{document}

\begin{abstract}
We study the resonant enhancement of linear and nonlinear chiroptical effects in asymmetric silicon metasurfaces supporting multipolar Mie resonances and quasi-bound states in the continuum (quasi-BICs). We demonstrate theoretically and observe in experiment the pronounced {\it linear circular dichroism} at the quasi-BIC resonances. We further find that both local field enhancement and third-harmonic signal are large for Mie resonances and some quasi-BIC modes. We explain the selectivity of the nonlinear enhancement by employing the concept of critical coupling being more favorable for the modes with moderately large radiative quality factors ($Q$ factors). We demonstrate experimentally strong nonlinear chiroptical response associated with high efficiency of the third-harmonic generation and large {\it nonlinear circular dichroism} varying from $+0.918\pm0.049$ to $-0.771\pm0.004$ for the samples with different asymmetries. We believe our results suggest a general strategy for engineering nonlinear chiroptical response in dielectric resonant metasurfaces. 
\end{abstract}

\newpage

{\it Chiral nanophotonics} is a rapidly expanding frontier of optics that has revolutionized the conventional approaches to the manipulation of the polarization states of light at the nanoscale~\cite{oh2015chiral,schaferling2017chiral}. Chiral systems represent a special class of objects that cannot be superimposed onto their mirror images~\cite{kelvin1894molecular}. This property allows for chiroptical effects such as {\it circular dichroism} (CD) - a difference in the outcomes of light-matter interaction that depends on chirality of the material and the direction of propagation of circularly polarized light. The chiral effects can be observed in both linear and nonlinear optical regimes~\cite{collins2019first,ohnoutek2022third,yuri2021}. The CD finds many applications in sensing, spectroscopy, negative refractive index engineering, with nanostructures made of chiral elements~\cite{xiong2010construction,valev2013chirality, mun2020electromagnetic}.  Over the last two decades many demonstrations of strong chiroptical effects were done in planar plasmonic~\cite{papakostas2003optical,ren2012giant,hentschel2017chiral} and dielectric ~\cite{solomon2018enantiospecific,tanaka2020chiral} metastructures, which avoid the fabrication complexity of bulky three-dimensional metamaterials. 

The practical use of planar plasmonic metasurfaces for chiral applications is hampered by low intrinsic chirality~\cite{papakostas2003optical,ren2012giant} which is generally prohibited for two-dimensional geometries because of their inherent mirror reflection symmetry. The engineering of an extrinsic chirality is one of the avenues for the enhancement of a three-dimensional chiral response in plasmonics limited by the oblique orientation of the pump direction with respect to the planar structure~\cite{plum2009extrinsic}. In the last few years, we observe the intense studies of strong intrinsic chiroptical effects in {\it dielectric metasurfaces} aimed to employ resonances associated with optically induced electric and magnetic Mie modes~\cite{wu2014spectrally,hu2017all,zhu2018giant,liu2020multipole,wang2021arbitrary, zhao2022realization}. Very recently, pronounced linear chiroptical response was predicted theoretically and demonstrated experimentally in chiral dielectric metasurfaces supporting sharp resonances associated with the physics of bound states in the continuum~\cite{gorkunov2020metasurfaces,overvig2021chiral,dionne2021,gorkunov2021bound,shi2022planar}.

{\it Nonlinear chiral metadevices} require a combination of strong chirality of nonlinear response and high efficiency of the frequency conversion making their development very challenging. Nonlinear chirality in plasmonic metamaterials attracted considerable attention in the past decade, however, the conversion generation efficiency is strongly limited by the absorption losses in metals~\cite{valev2009plasmonic,belardini2011circular,ren2012giant,rodrigues2014nonlinear,valev2014nonlinear,kolkowski2015octupolar,spreyer2022second}. High-index dielectric structures with large nonlinear susceptibilities and strong resonant field enhancement due to overlapped electric and magnetic Mie-resonant  modes represent a versatile platform for engineering strong optical response~\cite{kuznetsov2016optically,kivshar2018all}. In the recent years, nonlinear dielectric metasurfaces with engineered Mie resonances were developed for the generation of optical harmonics with the conversion efficiencies far much exceeding their plasmonic counterparts~\cite{yang2015nonlinear,tong2016enhanced,liu2018enhanced,wang2018nonlinear,vabishchevich2018enhanced,marino2019zero, fedotova2020second}. A very appealing idea is to employ these resonances for enhancing both linear and nonlinear chiroptical effects. Recently, it was demonstrated that a nanoparticle dimer made of AlGaAs provides high CD for the second-harmonic signal due to the excitation of Mie resonances~\cite{frizyuk2021nonlinear}. 

For optical metasurfaces, we may also employ collective resonances such as {\it bound states in the continuum}~(BICs) which represent special optical resonances with infinite values of the quality factor ($Q$ factor)~\cite{hsu2016bound, koshelev2022bound}. In practice, BICs are manifested as quasi-BICs with the high but finite values of the $Q$ factor~\cite{koshelev2020dielectric}. The quasi-BIC resonances provide a strong local field enhancement combined with the subwavelength confinement of light and tunability of metasurfaces~\cite{koshelev2018asymmetric}. Recent studies demonstrated that nonlinear dielectric metasurfaces supporting quasi-BICs can be employed for the enhanced harmonic generation and strong self-action effects~\cite{koshelev2019nonlinear,liu2019high,anthur2020continuous,sinev2021observation,zograf2022high}. 

However, in many cases the performance of nonlinear metasurfaces at the quasi-BIC resonances is limited due to fabrication-induced losses contributed by surface scattering, disorder, and finite-size effects~\cite{sadrieva2017transition}. The typical measured value of $Q$ factor at the quasi-BIC resonance in the visible and IR range for many dielectric metasurfaces varies between $100$ and $1000$~\cite{ha2018directional,cui2018multiple,tittl2018imaging,koshelev2019nonlinear,anthur2020continuous,sinev2021observation, jahani2021imaging,murai2022engineering,zograf2022high, shi2022planar}, depending on the fabrication method, setup limitations, and wafer quality~\cite{kuhne2021fabrication}. We notice that in some experimental studies the Q factor of the quasi-BIC resonances supported by dielectric metasurfaces exceeds $1000$ by the order of magnitude~\cite{liu2019high}, being a subject to a superior fabrication quality.  A mismatch between a weak coupling of light to the quasi-BIC metasurface and strong fabrication-induced losses leads to a decrease of the local field enhancement~\cite{koshelev2019nonlinear,bernhardt2020quasi}. As a result, for nonlinear applications the benefit of using quasi-BIC resonances over Mie resonances with a lower radiative $Q$ factor remains unclear despite recent  predictions~\cite{gandolfi2021near,shi2022planar}.

In this Letter, we study dielectric resonant metasurfaces for the enhancement of both linear and nonlinear three-dimensional chiroptical effects.  We fabricate a set of chiral nonlinear metasurfaces designed to support resonant Mie modes and quasi-BIC modes in the near-IR frequency range. These engineered metasurfaces are composed of broken-symmetry $L$-shaped silicon nanoparticles with asymmetry that changes their degree of chirality. First, we measure the linear transmittance selectively for left- and right-circularly polarized (LCP and RCP) light in the near-IR frequency range and show experimentally the enhancement of linear CD in the vicinity of quasi-BIC resonances. Employing the numerical analysis of the modes, we characterize the quasi-BICs and demonstrate that the metasurface also supports a Mie mode with the $Q$ factor of $50$ which does not contribute sharp features into the linear response. Using the coupled-mode theory, we find that for the Mie mode and some quasi-BICs the local field and nonlinear polarization enhancement are a few order of magnitude larger satisfying better the critical coupling condition than other quasi-BIC resonances.  Next, we demonstrate experimentally the large enhancement of nonlinear CD for the third-harmonic (TH) signal with a high generation efficiency at the resonance with the Mie and selected quasi-BICs. We reveal that while keeping the conversion efficiency high, the nonlinear CD can be changed gradually from the value $0.918\pm0.049$ to $-0.771\pm0.004$ for samples with different asymmetry parameters. We believe our results suggest a practical approach for engineering resonant nonlinear dielectric metasurfaces with strong chiroptical response.

\section{Results and Discussion}

\begin{figure}[t]
  \includegraphics[width=0.95\linewidth]{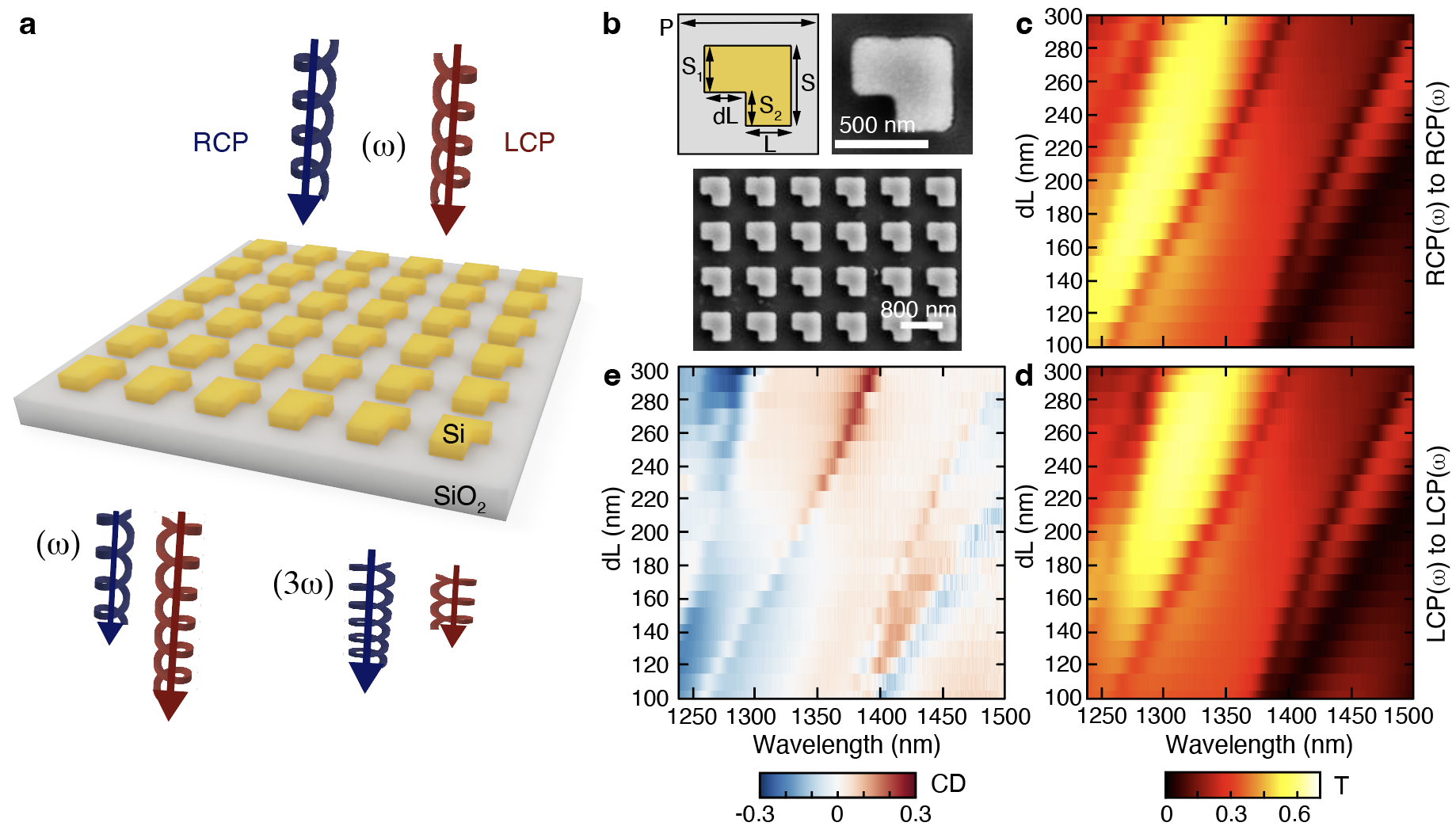}
  \caption{ Chiral metasurfaces and their linear characteristics. Experimental results. (a,b) Schematic and SEM images of a silicon metasurface with broken-symmetry $L$-shaped meta-atoms. (c,d) Experimentally measured co-polarized transmittance $I^{(\omega)}_{\rm RCP}$ for RCP (c) and $I^{(\omega)}_{\rm LCP}$ for LCP (d) excitation vs. pump wavelength and $dL$. (e) Measured CD of the transmitted signal vs. pump wavelength and $dL$. The experimental data are not interpolated.}
  \label{fig:1}
\end{figure}

Our metasurfaces consist of the $L$-shaped meta-atoms placed on top of a fused silica substrate arranged in a square lattice, as shown in Fig.~\ref{fig:1}a. The geometrical parameters of the unit cell are $P=840$~nm, $S=525$~nm, $S_1=315$~nm, $S_2=210$~nm, $L=315$~nm and the meta-atom height is $315$~nm, as shown in Fig.~\ref{fig:1}b. The material parameters are described in Supporting Information. We engineer the metasurface to support resonant modes in the near-infrared spectral range with radiative Q factors from a few tens to a few thousand. We fabricate a set of samples with the parameter $dL$ changing from $100$ to $300$~nm with the step of $10$~nm. The scanning electron microscope (SEM) image of the meta-atom and metasurface is shown in Fig.~\ref{fig:1}b. We note that because of the meta-atom shape and substrate presence, the eigenstates of the Jones matrix are elliptically polarized so the structure supports both linear and circular birefringence and dichroism~\cite{menzel2010advanced,kruk2015polarization}. While for the study of linear chiral response meta-atoms possessing a rotational symmetry may be more favourable, for nonlinear response the microscopic symmetry of nonlinearities dominates the effects driven by the symmetry (such as C$_4$) of meta-atoms and metasurfaces~\cite{frizyuk2021nonlinear,nikitina2022when}.

First, we characterize the linear chiroptical response of the fabricated metasurfaces by measuring experimentally co-polarized transmission $I^{(\omega)}_{\rm RCP}$ and $I^{(\omega)}_{\rm LCP}$. We use  selective excitation with a polarized white light and selective collection of the same polarization at the output. The maps of experimental transmission spectra are shown in Figs.~\ref{fig:1}(c,d). The spectra demonstrate sharp resonant features in the near-IR frequency range from $1240$ to $1500$~nm. We associate these spectral features with five distinct resonant modes supported by the metasurface. The linear CD is evaluated as $({I^{(\omega)}_{\rm RCP}-I^{(\omega)}_{\rm LCP}})/({I^{(\omega)}_{\rm RCP}+I^{(\omega)}_{\rm LCP}})$. The measured CD shows pronounced enhancement up to $0.254$ and $-0.300$ in the vicinity of the resonant modes, as shown in Fig.~\ref{fig:1}e. The signal is measured normal to the metasurface plane for the wavelengths above the diffraction threshold of $1225$~nm. We notice that the observed linear CD is driven by the interaction of the meta-atom resonant modes with the substrate because the meta-atoms are geometrically achiral~\cite{menzel2008retrieving,maslovski2009symmetry,gorkunov2021bound}. The details of the measurement setup, as well as maps of cross-polarized transmitted signal and asymmetric circular transmission are summarized in Supporting Information.

\begin{figure}[t]
  \includegraphics[width=0.75\linewidth]{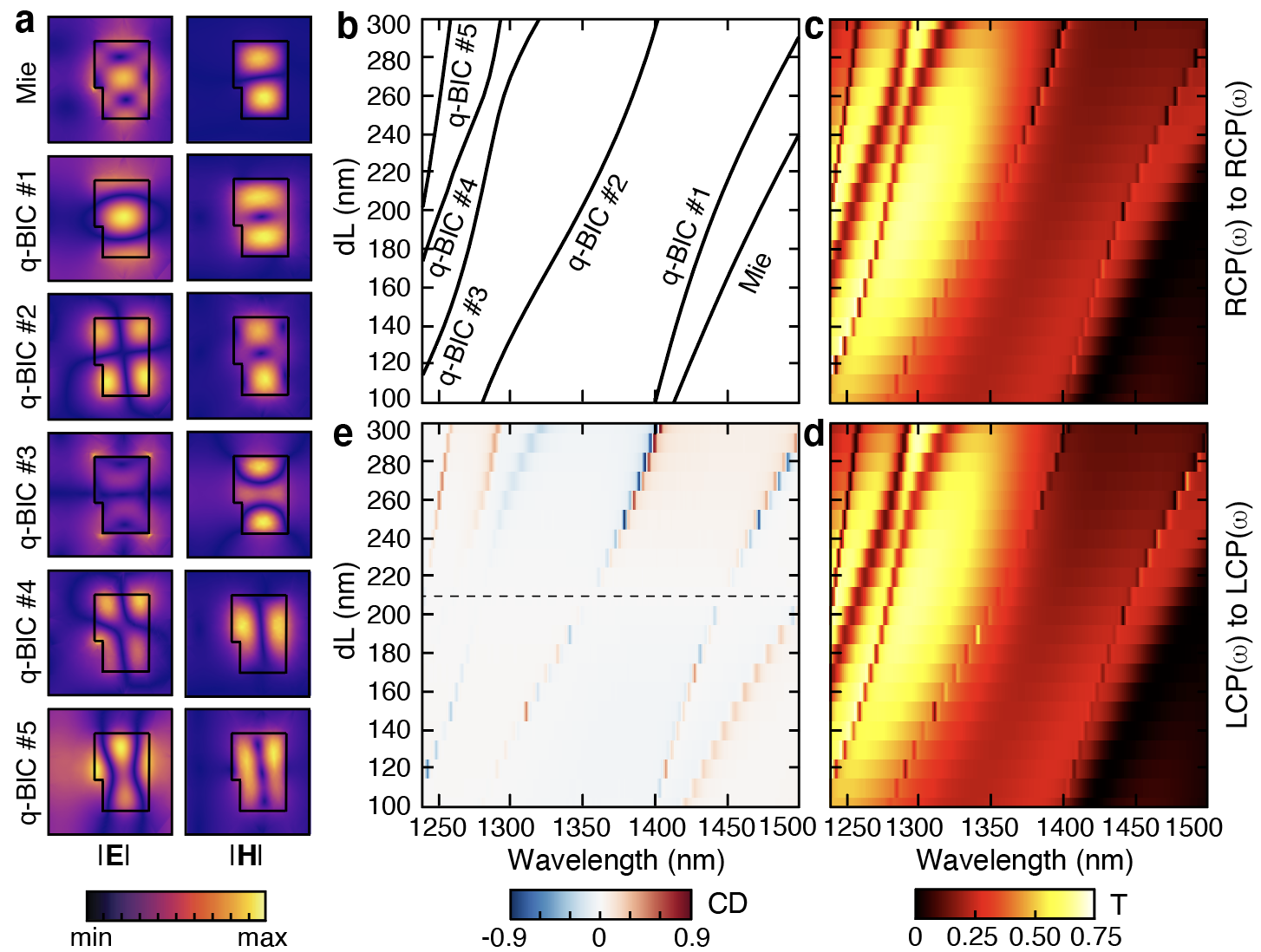}
  \caption{  Numerical results. (a) Near-field profiles of the metasurface resonant modes (marked as "Mie"  and "q-BIC", the latter stands for "quasi-BIC") in $1240$~nm to $1500$~nm wavelength range. (b) Mode resonant wavelengths vs. asymmetry parameter $dL$. (c-e) Co-polarized transmittance for RCP (c) and LCP (d) excitation vs. pump wavelength and $dL$. (e) Linear CD vs. pump wavelength and $dL$. The dashed line shows $dL=S_2=210$~nm sample with zero CD due to accidental symmetry. }
  \label{fig:2}
\end{figure}

Next, we analyse and characterize the resonant features observed in experiment by searching for metasurface eigenmodes for the asymmetry parameter $dL$ varying from $100$ to $300$~nm. The simulations reveal that the metasurface supports six resonant modes in the wavelength range from $1240$ to $1500$~nm, which we label, after analyzing their properties, as Mie and quasi-BICs $\#1$ to $\#5$. The near-field mode profiles and evolution of their resonant wavelengths are shown in Figs.~\ref{fig:2}a,b. The nature of Mie and quasi-BICs is confirmed by the simulation of Q factor for $dL=0$ for a free-standing metasurface showing the divergence for quasi-BICs (see Supporting Information). For comparison with experiment, we calculate numerically the co-polarized transmission and linear CD maps  summarized in Figs.~\ref{fig:2}c-e. We observe that the mode resonant wavelengths match the position of transmittance and CD features. The calculated linear CD reaches up to $+0.801$ and $-0.904$ at the resonances. The calculated linear CD for metasurfaces without a substrate vanishes, see Supporting Information. The simulation details and additional linear birefringence calculations are available in Supporting Information.

The experimental data shown in Figs.~\ref{fig:1}c-e are in a good agreement with our numerical results of Figs.~\ref{fig:2}c-e. The experimentally measured resonant features exhibit a blue shift by about $20$~nm and linewidth increase compared to the numerical data. From the experimental transmittance, we estimate that the total $Q$ factor of the resonant modes is limited by the value $Q_{\rm tot}=50$. The resonance broadening and smaller values of experimental CD can be explained by (i) fabrication imperfections, (ii) finite sample size, (iii) the high ${\rm NA}=0.25$ of the excitation objective. We estimate the fabrication-induced $Q$ factor $Q_{\rm fab}$ between $50$ and $300$ using $Q_{\rm fab}=Q_{\rm tot}$, as the lower estimate, and experimental data for similar Si metasurfaces, as the upper estimate~\cite{koshelev2019nonlinear, kuhne2021fabrication}. The effect of a focused beam excitation leads to the Q factor averaging in the reciprocal space accounting for the band curvature in the vicinity of the $\Gamma$ point. In further discussions, we omit this effect because we use a low-NA focusing lens for nonlinear measurements.

For strong chiroptical effects of the third-harmonic signal generated from Si metasurfaces, it is required to achieve the pronounced chiral response of the nonlinear signal combined with a high energy conversion efficiency. For resonant Si metasurfaces, the local field intensity at the TH wavelength $I_{\mathrm{loc}}^{(3\omega)}$ depends on the enhancement of the local field intensity $I_{\mathrm{loc}}^{(\omega)}$ as $I_{\mathrm{loc}}^{(3\omega)}\propto \left(I_{\mathrm{loc}}^{(\omega)}\right)^3$. We estimate the dependence of the enhancement on the rates of resonant losses as $I_{\mathrm{loc}}^{(\omega)}\propto {Q_{\rm tot}^2}/{Q_{\rm rad}}$ similar to plasmonic antennas~\cite{seok2011radiation}, where $Q_{\rm rad}$ is the mode radiative $Q$ factor. For dielectric structures, the absorption losses are negligible, and the nonlinear field intensity at the resonance can be presented in the form~\cite{koshelev2019nonlinear} 
\begin{equation}
    I_{\mathrm{loc}}^{(3\omega)}\propto Q_\text{fab}^3\left[\frac{Q_\text{rad}Q_\text{fab}}{(Q_\text{rad}+Q_\text{fab})^2}\right]^3.
    \label{eq:1}
\end{equation}
In the recent studies of quasi-BICs in nonlinear dielectric metasurfaces~\cite{koshelev2019nonlinear,bernhardt2020quasi}, it was shown that in analogy to absorptive structures, both $I_{\mathrm{loc}}^{(\omega)}$ and $I_{\mathrm{loc}}^{(3\omega)}$ are maximized by mimicking the so-called \textit{critical coupling condition}~\cite{yariv2002critical}
\begin{equation}
Q_\text{rad}=Q_\text{fab}.
\label{eq:2}
\end{equation}

\begin{figure}[t]
  \includegraphics[width=0.95\linewidth]{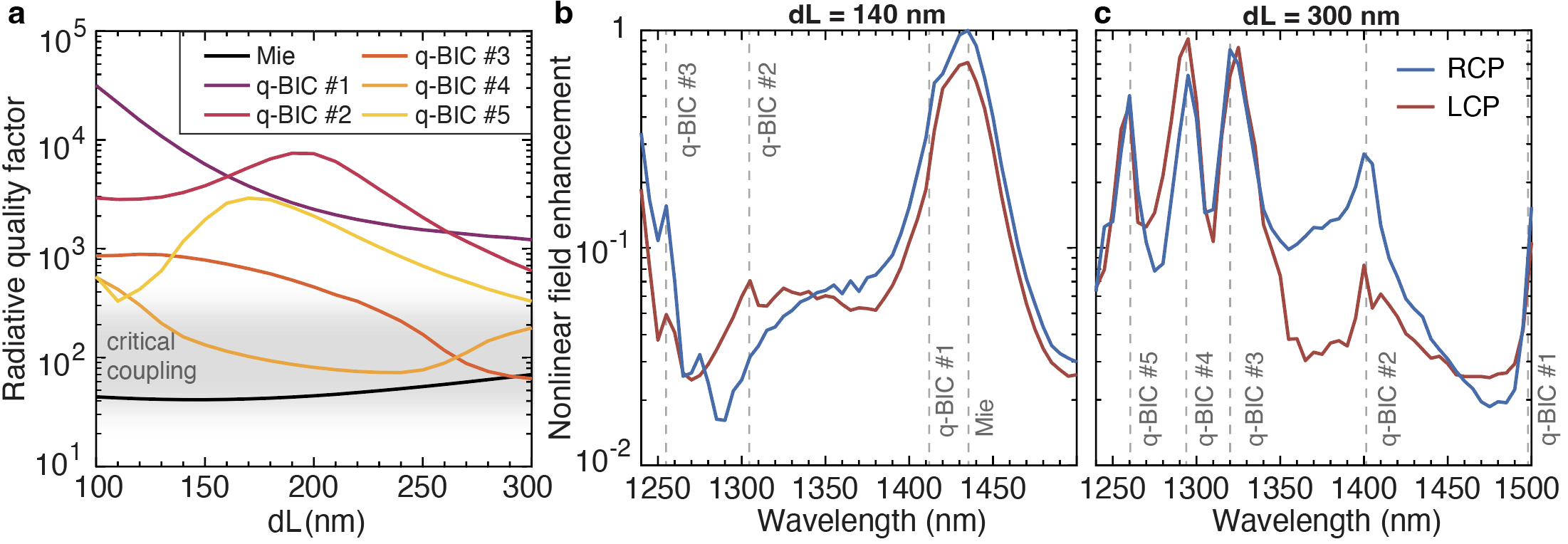}
  \caption{Numerical results. (a) Calculated $Q_{\rm rad}$; the critical coupling range is marked with a blurred gray area. (b,c) Simulated nonlinear local field enhancement for RCP (blue) and LCP (red) excitation for $dL=140$~nm (b) and $dL=300$~nm (c). Mode resonant wavelengths are marked with gray solid lines. Spectra in (b,c) are normalized on the maximum value for selected values of $dL$.}
  \label{fig:3}
\end{figure}

Figure~\ref{fig:3}a shows the dependence of $Q_{\rm rad}$ on the asymmetry parameter $dL$ for the resonant modes. The $Q$ factors of the Mie mode and quasi-BICs $\#3,4,5$ are within the critical coupling range from $50$ to $300$, shown with a gray blurred area. Figures~\ref{fig:3}b,c show the results of the full-wave numerical simulations of the nonlinear local field enhancement for RCP and LCP excitations for $dL$ varying from $140$~nm to $300$~nm, respectively. We select samples with a different $dL$ to highlight the effect of critical coupling. The calculations confirm that the Mie and some quasi-BIC modes provide much stronger field enhancement than other quasi-BICs $\#1,2$, in agreement with Eqs.~(\ref{eq:1}) and (\ref{eq:2}).

We further simulate and measure a TH intensity for the set of designed Si metasurfaces and evaluate the CD of the nonlinear signal. For nonlinear measurements, we pump the metasurface from the top (air side) with a spectrally tunable femtosecond laser and collect the generated TH signal in the transmission (forward) direction in the zero diffraction order with a spectrometer equipped with an EMCCD detector. We evaluate the intensities of forward scattered TH signal $I^{(3\omega)}_{\rm RCP}$ and $I^{(3\omega)}_{\rm LCP}$ for co-polarized RCP (and LCP) pump and harmonic, respectively. We evaluate the nonlinear CD for the TH signal as $({I^{(3\omega)}_{\rm RCP}-I^{(3\omega)}_{\rm LCP}})/({I^{(3\omega)}_{\rm RCP}+I^{(3\omega)}_{\rm LCP}}$). For nonlinear simulations, we use the undepleted pump approximation and model the effect of fabrication-induced losses by adding an auxiliary extinction coefficient corresponding to $Q_{\rm abs} \simeq 100$. The TH signal is calculated in transmission (forward) direction in the zero diffraction order. The details of numerical simulations and experimental data are available in Supporting Information.

\begin{figure}[t]
  \includegraphics[width=0.85\linewidth]{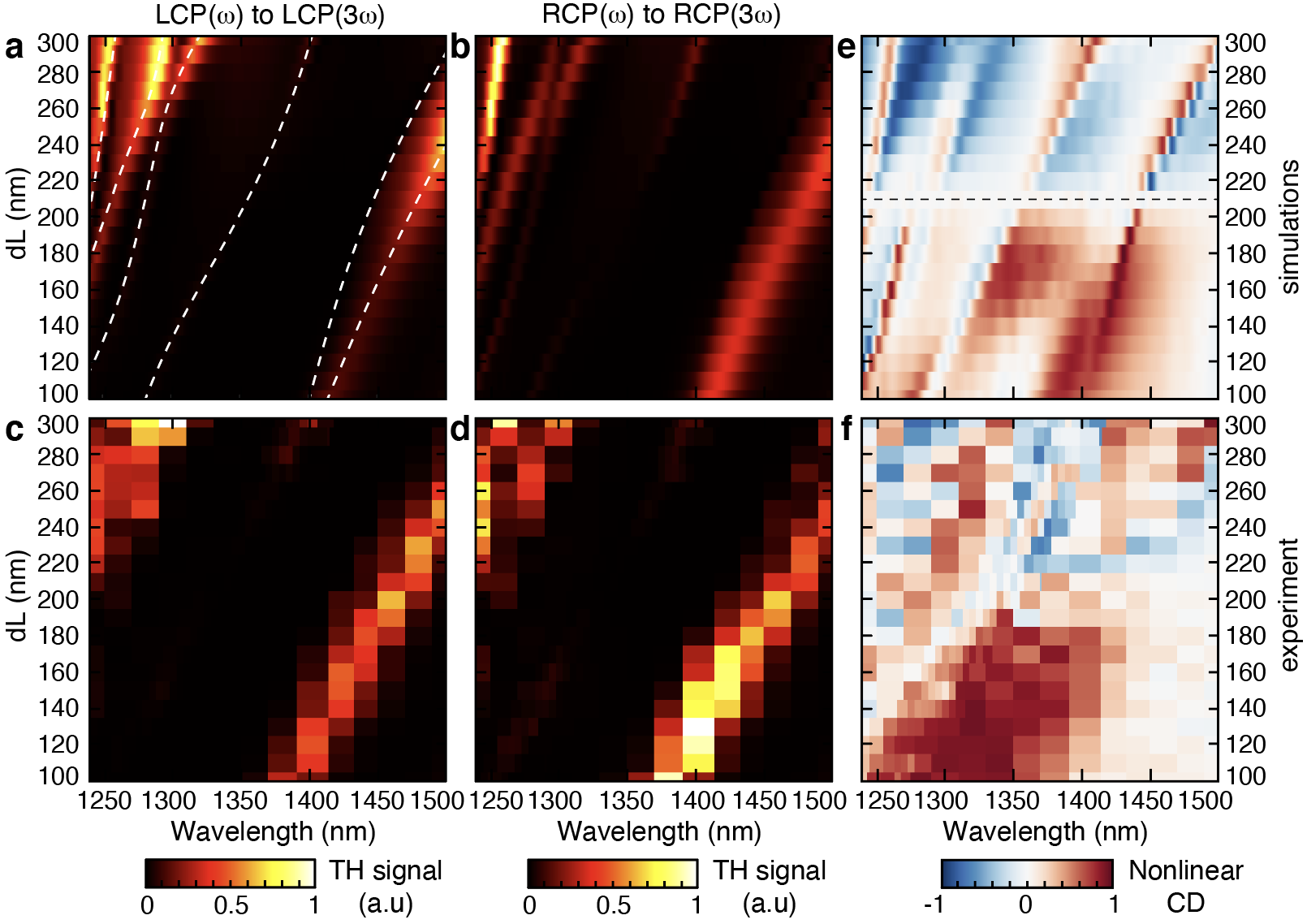}
  \caption{Experiment vs. theory. Numerical (a,b) and experimental (c,d) co-polarized TH intensities $I^{(3\omega)}_{\rm LCP}$ for LCP (a,c) and $I^{(3\omega)}_{\rm RCP}$ RCP (b,d) excitation vs. pump wavelength and $dL$. The mode resonant wavelengths are marked with white dashed lines in (a). (e,f) Simulated (e) and measured (f) nonlinear CD of the TH signal vs. pump wavelength and $dL$. The dashed line shows $dL=S_2=210$~nm sample with zero CD due to accidental symmetry. The spectra in (a-d) are independently normalized for simulations and experiment on the maximum measured value for all samples. The experimental data are not interpolated.}
  \label{fig:4}
\end{figure}

Figures~\ref{fig:4}a,b show the simulated TH signal that is enhanced along the dispersion curves of the Mie and quasi-BIC $\#3,4,5$ modes at the pump wavelength, in agreement with Figs.~\ref{fig:3}b,c and the critical coupling model. For quasi-BICs $\#1,2$, the TH signal enhancement is generally more than an order of magnitude weaker with higher values for large $dL$ explained by the decrease of $Q_{\rm rad}$ (see Fig.~\ref{fig:3}a). The simulated TH CD is resonantly enhanced in the vicinity of all six modes, with the value up to $+0.990$ and $-0.974$, as shown in Fig.~\ref{fig:4}e. The nature of observed nonlinear chirality is connected with the combination of the microscopic symmetry of nonlinearities and the symmetry of the meta-atom and lattice~\cite{frizyuk2021nonlinear,nikitina2022when}. We notice that the calculated linear and nonlinear CD vanishes for $dL=210$~nm because of an accidental in-plane mirror symmetry of the meta-atom, see a black dashed line in Fig.~\ref{fig:2}e and Fig.\ref{fig:4}e.  The TH signal calculations without a substrate show that the nonlinear CD is nonzero, despite the linear CD vanishes, see Supporting Information. 

The measured TH intensities and CD are in a qualitative agreement with the numerical results, as shown in Figs.~\ref{fig:4}c,d,f. The maxima of TH intensities are blue shifted by about $20$~nm with respect to the numerical results, as for the case of the linear measurements. The resonant wavelengths extracted from the experimental transmittance and projected on the measured TH maps show that the TH enhancement is driven by the resonances at the pump wavelength, in agreement with the numerical results, see Supporting Information. The experimental data show clear signature of Mie and  quasi-BIC $\#3,4,5$ modes while the contribution of quasi-BICs $\#1,2$ is visible only for large values of $dL$. Figure~\ref{fig:4}f shows the spectra of measured TH CD reaching large values up to $+0.918\pm0.049$ and $-0.771\pm0.004$. The differences between the numerical and experimental CD TH explained by a significant measurement error for low TH intensities.

We notice that for nonlinear resonant nanostructures the THG efficiency can also depend on the mode structure at the harmonic frequency~\cite{celebrano2015mode,koshelev2020subwavelength}. In our case, the density of modes at the TH wavelength is high, and they overlap strongly because of a low Q factor at the TH wavelength (limited by $4$ at $500$~nm and $1$ at $415$~nm) due to high absorption losses. Thus, we consider the structure as nonresonant at the TH wavelength, and this is confirmed by TH directionality simulations (see Supporting Information). The TH intensity maps for cross-polarized excitation and collection, and the nonlinear CD spectra evaluated using the alternative definitions are presented in  Supporting Information.

\begin{figure}[t]
  \includegraphics[width=0.85\linewidth]{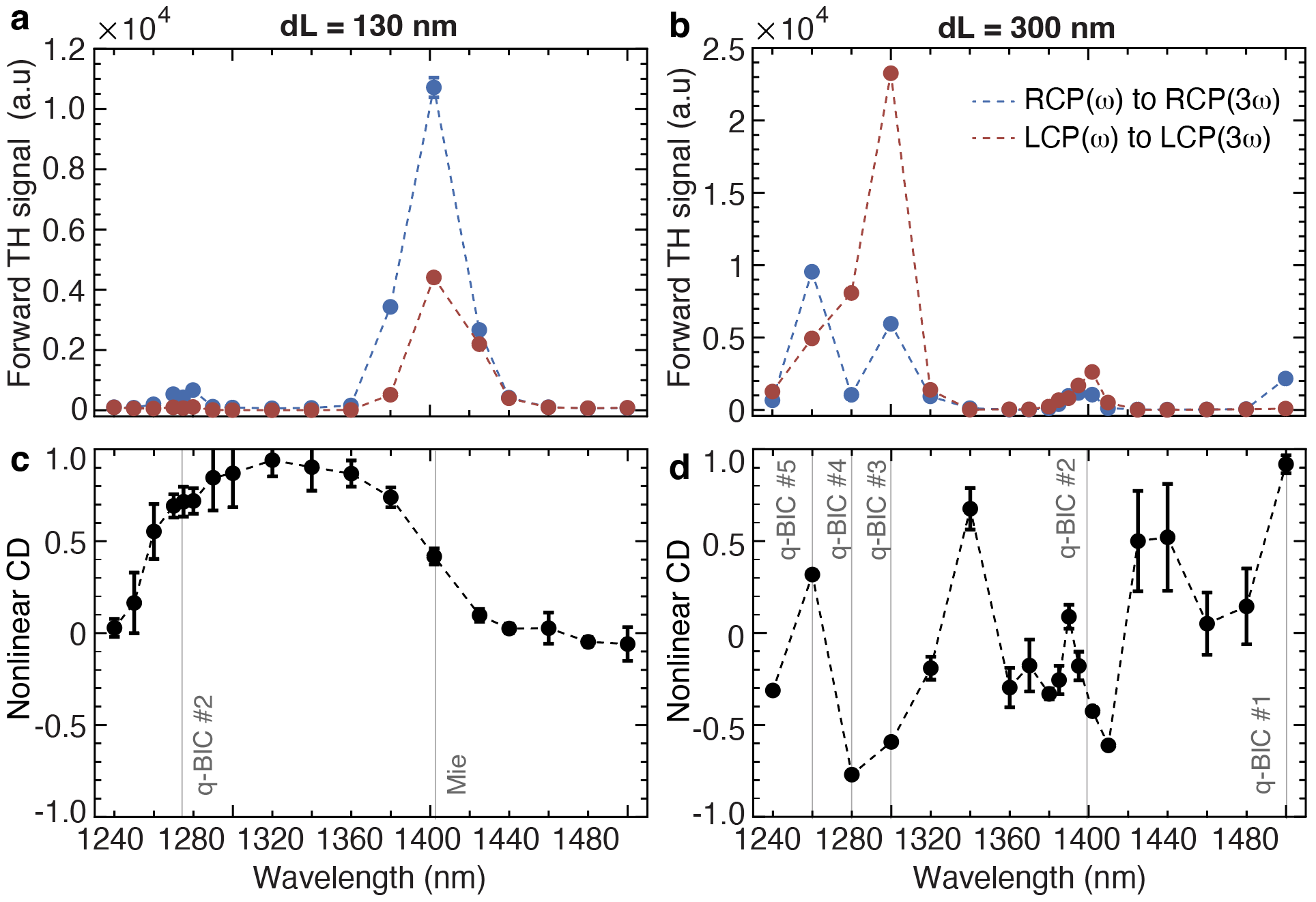}
  \caption{Experimental results. (a,b) Measured forward TH signal for co-polarized RCP (blue) and LCP (red) excitation and collection for $dL=130$~nm (a) and $300$~nm (b). (c,d) Measured TH CD spectrum for $dL=130$~nm (c) and $300$~nm (d). Dashed lines are guide for eyes. The resonant wavelengths are marked with solid vertical lines.}
  \label{fig:5}
\end{figure}

As discussed above, a combination of THG efficiency and large nonlinear CD is required for achieving strong nonlinear chiroptical response. The results shown in Fig.~\ref{fig:4} demonstrate that this combination can be achieved for resonant dielectric metasurfaces in the vicinity of resonances. We study in detail the measured TH signal and nonlinear CD spectra for the samples with $dL=130$~nm and $dL=300$~nm, shown in Fig.~\ref{fig:5}, providing the best nonlinear performance. For $dL=130$~nm, TH is maximized at $1402$~nm pump wavelength, which corresponds to the excitation of the Mie mode and yields  TH CD of $+0.417\pm0.044$. Another increase of the TH intensity is realized at the quasi-BIC $\#2$ wavelength of $1275$~nm with TH CD of $0.719\pm0.081$. For a larger sample with $dL=300$~nm, the TH intensity demonstrates increased values at $1260$, $1280$, $1300$, $1402$, and $1500$~nm with TH CD of $+0.318\pm0.022$, $-0.771\pm0.004$, $-0.593\pm0.002$, $-0.425\pm0.011$, and $+0.918\pm0.049$, respectively. These peaks correspond to the excitation of quasi-BICs $\#5,4,3,2,1$, respectively, in a direct agreement with the critical coupling concept and Fig.~\ref{fig:3}c. The CD maps shown in Figs.~\ref{fig:4}e,f reveal that TH CD can be gradually changed from a large negative value to a large positive value with a change of the asymmetry parameter $dL$ and pump wavelength along the dispersion of resonant modes.



In conclusion, we have studied linear and nonlinear chiral response of dielectric resonant metasurfaces. We have designed and fabricated a set of silicon metasurfaces composed of broken-symmetry $L$-shaped meta-atoms supporting Mie and quasi-BIC resonant modes in the frequency range of $1240$~nm to $1500$~nm. As expected, we have observed the pronounced enhancement of linear CD in the vicinity of quasi-BIC resonances with the magnitude reaching up to $0.3$. This value is comparable with the values from $0.1$ to $0.97$ demonstrated in the near-IR range with earlier implementations using other approaches for plasmonic and dielectric metasurfaces~\cite{hentschel2012optical,cui2014giant,chen2018chiral,wu2018high,semnani2020spin,zhu2018giant,ji2021artificial,shi2022planar}. However, the enhanced nonlinear CD has been observed for the Mie mode and a few quasi-BIC modes with the nonlinear conversion efficiency exceeding the THG efficiency at other quasi-BIC resonances by at least one order of magnitude. We have explained this observation by the critical coupling condition required for maximizing the nonlinear local field enhancement in resonant metasurfaces with losses. We have demonstrated that nonlinear CD can be varied from large positive values (more than $+0.9$) to large negative values (below $-0.75$) for samples with different asymmetries. The observed nonlinear CD exceeds the best-to-date values achieved for chiral nonlinear dielectric metasurfaces and oligomers~\cite{frizyuk2021nonlinear,shi2022planar}. Compared to the earlier studies of chiral nonlinear plasmonic metasurfaces showing strong nonlinear CD, the achieved harmonic generation efficiency is sufficiently larger~\cite{valev2009plasmonic,belardini2011circular,ren2012giant,rodrigues2014nonlinear,valev2014nonlinear,kolkowski2015octupolar,spreyer2022second}. Our results demonstrate how to achieve strong nonlinear chiroptical effects in resonant dielectric metasurfaces with the maximization of frequency conversion efficiency.

\begin{acknowledgement}

The authors acknowledge numerous highlighting comments from  M.~Gorkunov and V.~Valev, and a help of Ms. Mingke Jin with the SEM measurements. The sample fabrication was conducted as part of a user project at the Center for Nanophase Materials Sciences (CNMS), which is a US Department of Energy, Office of Science User Facility at Oak Ridge National Laboratory. G.L. was supported by National Natural Science Foundation of China (grants 91950114 and 12161141010). Y.K. acknowledges a support from the Australian Research Council (grant DP210101292), as well as the International Technology Center Indo-Pacific (ITC IPAC) and Army Research Office under Contract No. FA520921P0034.

\end{acknowledgement}

\begin{suppinfo}

The Supporting Information is available free of charge at [?].
Materials and methods, including details on numerical methods, sample fabrication, optical measurements. Numerical calculations, including transmittance, birefringence and dichroism for linearly polarized pump; transmittance, TH signal, CD and TH CD for a free-standing metasurface; directionality proprieties of TH signal; eigenmode analysis describing the nature of resonant modes. Experimental data, including cross-polarized transmittance and TH signal, asymmetric transmission; raw spectra for transmittance, TH signal, CD and TH CD for all samples; mode analysis for experimental data.

\end{suppinfo}

\bibliography{References.bib}

\end{document}